\documentclass{aa}

\usepackage{graphicx}
\usepackage{hyperref}
\usepackage{natbib}                 
\bibpunct{}{}{;}{a}{}{,}
\usepackage{amsmath}                
\usepackage{txfonts}
\usepackage{lscape}

\usepackage{times}
    \usepackage{color}  

\newcommand{\ltsima} {$\; \buildrel < \over \sim \;$}
\newcommand{\gtsima} {$\; \buildrel > \over \sim \;$}
\newcommand{\lta} {\lower.5ex\hbox{\ltsima}}
\newcommand{\gta} {\lower.5ex\hbox{\gtsima}}

\begin{document}
   \title{XMM-Newton and Chandra X-ray follow-up observations of the VHE gamma-ray source HESS J1507-622}


{\small
\author{O.~Tibolla \inst{1}  
 \and S.~Kaufmann \inst{2}
 \and K.~Kosack \inst{3}
}
}

\institute{\small
ITPA, Universit\"at W\"urzburg, Campus Hubland Nord, Emil-Fischer-Str. 31 D-97074 W\"urzburg, Germany.
\and
Landessternwarte, Universit\"at Heidelberg, K\"onigstuhl 12, D-69117 Heidelberg, Germany
\and
IRFU/DSM/CEA, CEA Saclay, F-91191
Gif-sur-Yvette, Cedex, France
}

\offprints{O.~Tibolla\\
  \email{omar.tibolla@gmail.com ; Omar.Tibolla@astro.uni-wuerzburg.de}}

   \date{Received -- -- --; accepted -- -- --}

 
  \abstract
{The discovery of the unique source HESS J1507-622 in the very high energy (VHE) range (100 GeV-100 TeV) opened new possibilities to study the parent population
of ultra-relativistic particles found in astrophysical sources and underlined the possibility of new scenarios/mechanisms crucial for understanding the underlying astrophysical processes
in nonthermal sources.}
{The follow-up X-ray (0.2 - 10 keV) observations on HESS J1507-622 are
  reported, and possibilities regarding the nature of the VHE source
  and that of the newly discovered X-ray sources are
  investigated.}
{We obtained bservations with the X-ray satellites \emph{XMM-Newton} and
  \emph{Chandra}. Background corrections were
    applied to the data to search for extended diffuse
  emission. Since HESS J1507-622 covers a large part of the field of
  view of these instruments, blank-sky background fields were
  used.  }
{The discovery of several new X-ray sources and a new, faint, extended
  X-ray source with a flux of $\sim 6 \times 10^{-14} \; \rm{erg \;
    cm^{-2} \; s^{-1}}$ is reported. Interestingly, a new, variable
  point-like X-ray source with a flux of $\sim 8 \times 10^{-14} \;
  \rm{erg \; cm^{-2} \; s^{-1}}$ appeared in the 2011 observation,
  which was not detected in the previous X-ray observations.}
{The X-ray observations revealed a faint, extended X-ray source
  that may be a possible counterpart for HESS J1507-622. This
  source could be an X-ray pulsar wind nebula (PWN) remnant of the larger gamma-ray
  PWN, which is still bright in IC emission.
Several interpretations are proposed to explain the newly detected variable X-ray source.
}

   \authorrunning{O. Tibolla et al.}
   \titlerunning{HESS J1507-622 X-ray follow-ups}
   \maketitle
%

\section{Introduction}


Very high energy (VHE, $> 10^{11}$ eV) particles can be traced within
our Galaxy by a combination of nonthermal X-ray emission and gamma-ray
emission produced via leptonic (e.g.  inverse-Compton
scattering of electrons, Bremsstrahlung or synchrotron radiation) or
hadronic processes (decay of charged and neutral pions, interactions of
energetic hadrons with target material).  X-ray observations
of Galactic VHE sources are particularly important when investigating the nature of these sources, since they sample the synchrotron emission spectrum and therefore provide a way to distinguish between leptonic and hadronic models of gamma-ray emission.  The importance of this approach becomes immediately
evident, for example, for an uncooled electron
population in the interstellar medium (ISM) magnetic field: in
first approximation, the energy flux at TeV energies generated by inverse-Compton emission would be the same as the flux that we
would encounter at keV from synchrotron emsission.  Hence X-ray emission is expected from leptonic gamma-ray sources, and thus X-ray follow-up observations can be crucial for identifying VHE gamma-ray sources.

The discovery of HESS J1507-622 (\cite{1507}) challenged the standard
emission scenarios that have traditionally been associated with VHE sources and
opened the possibility of new scenarios in the high-energy astrophyics
field. HESS J1507-622 is among the brightest ($\sim$8\% of Crab flux)
sources in the extension of H.E.S.S. survey of the Galactic plane
(\cite{survey2}). But the newly discovered source lacks plausible
counterparts in the lower energy bands, as is the case for HESS
J1708-410 (\cite{unid}) and HESS J1616-508 (\cite{survey2}); but its
location is unique: while all unidentified VHE sources that
have been discovered in the H.E.S.S. Galactic Plane Survey so far are
located within $\pm$1 degree from the Galactic equator, HESS J1507-622
lies $\sim$3.5$^\circ$ from the Galactic plane. Considering the
comparably low hydrogen column density $n_H$ at 3.5$^{\circ}$ off the
plane, and hence the much lower Galactic absorption in X-rays
and the reduced background emission, one would expect to detect a
bright counterpart despite the anticipated lower spatial source
density of Galactic counterparts, but surprisingly, this is not the
case.

Moreover, since most Galactic VHE emitters are connected to young
stellar populations, which usually are concentrated near the Galactic disk, it
is unique to find an unidentified VHE gamma-ray source with a
3.5$^\circ$ offset from the Galactic plane. Our line-of-sight towards
the source direction intersects the Galactic disk with a scale height
of 50 pc up to a distance of at most 1 kpc extending farther into the Galactic halo. 
Both hadronic and leptonic scenarios were investigated to explain the VHE emission (\cite{1507}), and the latter, namely the ancient Pulsar Wind Nebulae (PWN) scenario (\cite{fd08}; \cite{d08}; \cite{myicrc}; \cite{gamma12}), seemed favored (\cite{1507}). This thesis seems strengthened by the recent \emph{Fermi LAT} detection (\cite{2FGL}; \cite{myicrc}). 
As a result of its small angular extension, the leptonic interpretation of HESS J1507-622 underlined the
uniqueness of this source, constraining this unidentified VHE source
to be located at a distance of $> 6$ kpc (as confirmed by a new time-dependent spatially-independent PWN model, \cite{Vost}, by the so-called classical ancient PWN model, \cite{myicrc} and by a leaky-box model, \cite{gamma12}) with an age of ($> 2 \times 10^4$ years; \cite{myicrc}), more distant and older than previously detected VHE PWNe.

In this article we present the several X-ray follow-up observations performed on this intriguing source with the current generation of X-ray observatories ({\it XMM-Newton} and {\it Chandra}).

\begin{figure*}[ht!]
\centerline{
\includegraphics[width=0.45\textwidth]{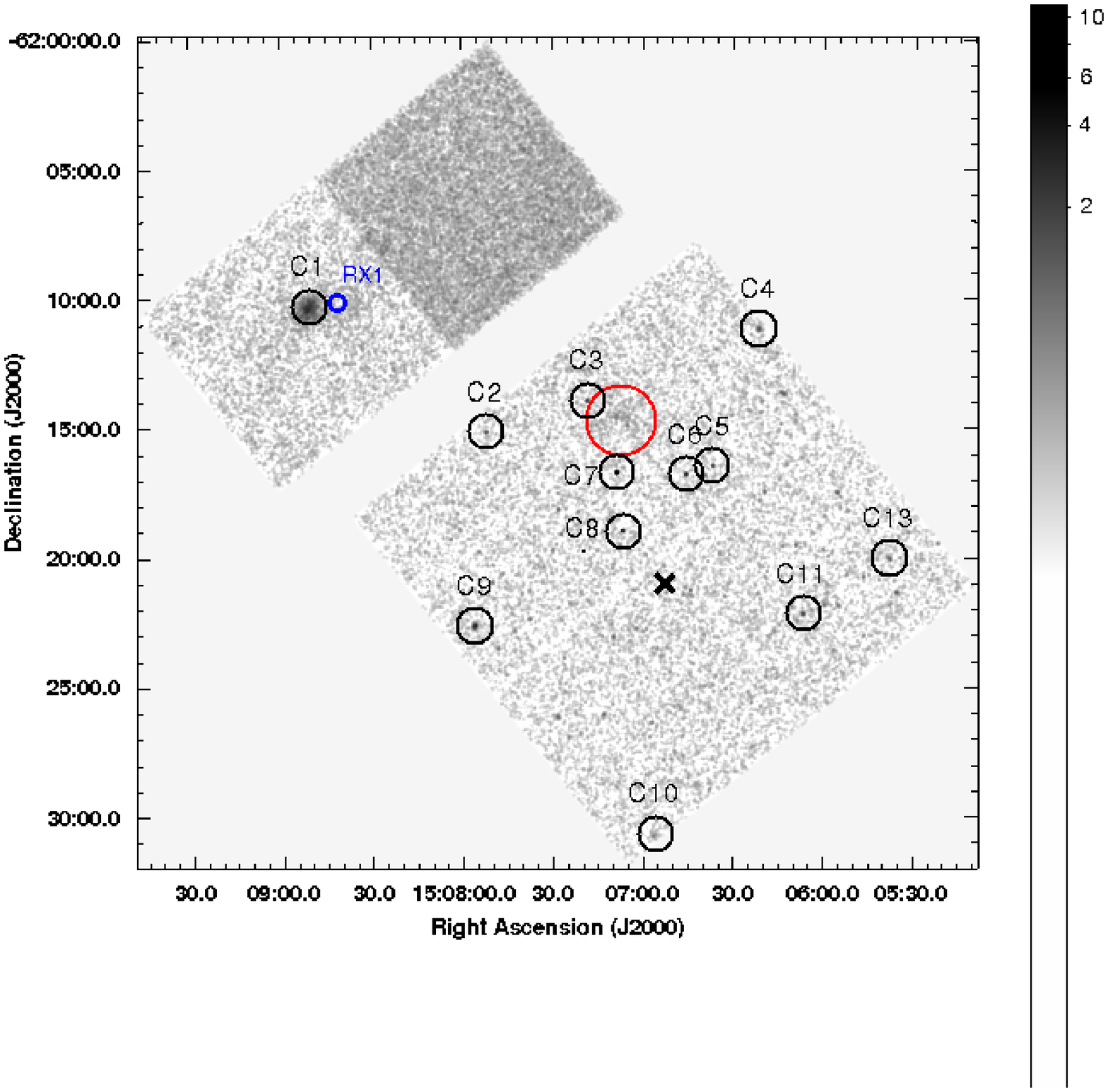}
\hfil
\includegraphics[width=0.45\textwidth]{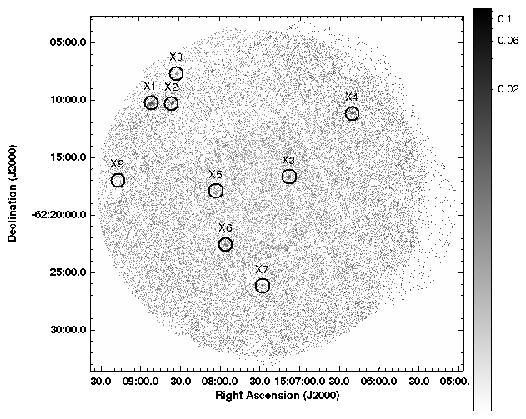}
}
\caption{{\it Left}:
smoothed and background subracted count map of the {\it Chandra} observation of HESS J1507-622. Black circles indicate the 12 sources detected by {\tt celldetect}. The blue circle indicates 1RXS J150841.2-621006, centered on its nominal position and with a radius corresponding to its positional uncertainities (\cite{faintRASS}). The faint extended emission described in the text is shown in red. The black cross indicates the centroid position of HESS J1507-622.
{\it Right}: mosaic of the MOS and PN detector images of the {\it XMM-Newton} observation in 2011. The single images for the three detectors have been exposure corrected. The circles represent the detected sources in the field of view (see also Table 1).
}
\label{chandra}
\end{figure*}


\section{X-ray observations and data analysis}

To attempt to detect an X-ray counterpart to HESS J1507-622, three exposures were made in the region around the source centroid with the current generation of X-ray satellites {\it XMM-Newton} and {\it Chandra}. 
Because of its offset of $\sim 3.5^\circ$ from the Galactic plane, the
absorption at the source location is approximately one order of
magnitude lower than for a source with no offset, which means that the column density at the position of HESS J1507-622 is $N_{\rm{H}} = 4.4 \times 10^{21} \;\rm{cm^{-2}}$ (LAB Survey, \cite{Kalberla2005}). 
Therefore HESS J1507-622 represents an excellent astrophysical laboratory for studying the unidentified VHE sources that so far have no obvious longer-wavelength counterpart (often referred to as ''dark accelerators``, \cite{unid}).

The first observations took place in 2009, when {\it XMM-Newton}
observed the source for 28 ks, and {\it Chandra} 20ks. Another {\it
XMM-Newton} observation was conducted in 2011 with an exposure of 39
ks, since the first observation was highly contaminated by soft proton
flares, which rendered large fraction of the dataset unusable.
Here we investigate these X-ray observations with particular attention to the newest {\it XMM-Newton} observation and to the X-ray sources that might be connected with the VHE gamma-ray source.


\subsection{{\it XMM-Newton}}

{\it XMM-Newton} observations were performed on January 27, 2009 and March 2, 2011 with 28 ks and 39 ks exposures, respectively. The {\it XMM-Newton} data were reprocessed with the most recent calibration information; the 2009 data were analyzed with {\tt SAS 9.0} and the 2011 data with {\tt SAS 11.0}. 
High-energy light curves ($>$10 keV for MOS and $10-12$ keV for PN)
were extracted to identify time periods affected by soft proton
flares. Both observations were affected by soft proton flares. The
cuts for the high energy light curve to identify the good time
interval (GTI) were $0.35$ counts/sec for MOS and $0.4$ counts/sec (as
recommended by the {\it XMM-Newton} team). To improve the data
analysis, GTIs with durations shorter than 100 sec were also removed,
since they mostly appeared during the proton flare periods. 

The PN and MOS2 detectors were used to search for sources in the field of view of {\it XMM-Newton} during these observations; the MOS1 detector was not considered since CCD6 has been switched off since March 2005 because of a micrometeorite impact\footnote{http://xmm.esac.esa.int/external/xmm\_news/items/MOS1-CCD6/}. The source detection algorithm {\tt edetect\_chain}, which uses sliding boxes, was applied simultaneously on images in five different energy ranges (0.1-0.5, 0.5-1, 1-2, 2-4.5, 4.5-12 keV) for the MOS2 and PN detector. The conversion factors for the medium filter mentioned in the 2XMM catalog user guide \footnote{http://xmmssc-www.star.le.ac.uk/Catalogue/2XMM/ \\ /UserGuide\_xmmcat.html} were used to determine the flux for each source.
The tool {\tt ewavelet}, which is better for detecting faint sources, was used on the MOS2 detector image, employing the whole energy range and taking into account the exposure map. The threshold for detection was chosen to be 5 sigma.

Radial profiles were determined using the tool {\tt eradial} to identify the extension of selected sources. With this tool, a King profile (representative of the point spread function, PSF) for a monochromatic energy was fit to the radial profile. 

To search for extended diffuse emission, the background from
astrophysical emission and instrumental noise must be
subtracted. Since the extension of the TeV source covers a large part
of the field of view, blank-sky background
fields\footnote{http://xmm2.esac.esa.int/external/xmm\_sw\_cal/background/
  \\ /blank\_sky.shtml} were used to estimate the astrophysical
background. With the {\tt BGSelector} tool, blank-sky observations
with the same detector, filter, observation mode, and values of
Galactic absorption ($4e21 < N_{\rm{H}} < 6e21\; \rm{cm^{-2}}$) were
selected. Backgrounds using blank-sky regions within a search radius of close to 4000 arcmin were tested as well, but lead to a
large number of regions with high Galactic absorption and thus too
large an increase of the background emission. Since the source is 3.5 degrees
away from the Galactic plane, a background region with similar
$N_{\rm{H}}$ is more appropriate. Instrumental background noise was
taken into account using the observations with a closed filter
wheel. The resulting background sky maps were reprojected onto the
observed coordinates and subtracted from the observation data set.

Images were created for the previously defined GTI for the PN and the two MOS detectors. The {\tt eexpmap} tool was used to obtain the exposure maps for each instrument, and {\tt mosaic} was used to sum them. The resulting image is shown in Fig. \ref{chandra}. 

The X-ray spectra were binned with the {\tt grppha} tool to obtain at
least 25 counts per bin (unless otherwise stated) to reach the
necessary statistical significance for $\chi^2$ statistics. The fit was obtained with  {\tt xspec}. For the spectral analysis, the energy ranges $0.1 - 10 \; \rm{keV}$ and $0.2 - 15 \; \rm{keV}$ were used for MOS and PN, respectively.

The optical monitor (OM, \citep{Mason2001}) onboard {\it XMM-Newton} observed the source in the filters (central wavelength) UVW2 (212 nm), UVM2 (231 nm), UVW1 (291 nm), U (344 nm) B (450 nm), and V (543 nm) simultaneously with the X-ray telescope. The analysis of these data was performed with the tool {\tt omichain}.

\subsection{{\it Chandra}}

{\it Chandra} observed HESS J1507-622 on June 6, 2009 with a total exposure of 20 ks. The {\it Chandra} data were analyzed using CIAO 4.1 and the calibration database CALDB v4.1.3. 
Since HESS J1507-622, with its extension of $0.15^{\circ}$, covers the whole ACIS-I field of view, the background 
was estimated using the standard blank-sky field files
 to search for faint and large extended emission. 
These blank-sky files were found using the {\tt acis\_bkgrnd\_lookup} tool and were weighted by the exposure of each observation. 
The background-corrected image is shown in Fig. \ref{chandra}. 

The source detection algorithm {\tt celldetect}, sensitive for point sources above a threshold of 3 sigma, was applied on the observed image, taking into account the exposure maps for each ccd.
The algorithm {\tt vtpdetect}, which is more sensitive for faint and extended sources, was also used to search for additional sources. Extensions (3 sigma) are given with the tool {\tt vtpdetect}.
The details of the sources found with these tools were summarized in Table 1 in \cite{1507}.

Light curves were produced with the tool {\tt dmextract} to search for variability. 

Radial profiles were extracted using circular annuli 
up to a radius of 100 pixel with the {\tt dmextract} tool. For these profiles, the background was detemined from an annular region with radius 100 pixel and outer radius 125 pixel centered on the source position. 

To check for source extension, we first estimated the PSF of the insturment for the given observations.  
The PSF for the specific on or off-axis angle was simulated using the Chandra Ray Tracer (ChaRT \footnote{http://asc.harvard.edu/chart/}), which simulates the High Resolution Mirror Assembly (HRMA) based on an input energy spectrum of the source and the exposure of the observation. With the software MARX \footnote{http://space.mit.edu/CXC/MARX/}, the output from ChaRT can be modeled taking into account instrument effects of the various detectors.
Unfortunately, this method cannot be used for the off-axis angle source on the detector ACIS-S, since, as explained in ChaRT issues and caveats\footnote{http://cxc.harvard.edu/chart/caveats.html}, MARX has problems projecting the rays if the source is located on one detector and the aim-point on the other, for example in this case the aim-point is on ACIS-I and the source is detected with ACIS-S. 
Therefore, the Chandra-provided monochromatic PSF (with the tool {\tt mkpsf}) for the energy closest to the peak energy of the observation was used.
The PSF was created for the energy 1.497 keV, which is one of the energies for which a PSF has been simulated in the PSF library. The offset angle of the source was taken into account. 
The PSF was scaled with the total number of source counts, and for extended sources this yields a higher surface brightness of the PSF at the center of the source that represents the profile of a point source.

The spectra of the detected sources were extracted from appropriate regions 
for the source and background, as described in the following sections. The redistribution matrix file (RMF) and ancillary response file (ARF) were calculated for each spectrum using the tools {\tt mkrmf} and {\tt mkarf}. 
The spectra were binned with at least 25 counts per bin (and 15 for the faint sources), and {\tt xspec v12.5} was used for the spectral analysis.


\section{Results of the data analysis}

\subsection{{\it XMM-Newton} observation in 2009}

The 28 ks observation with {\it XMM-Newton} in 2009 were partially reported in \cite{1507}, but are shown here in more detail to complete the X-ray coverage of the region of HESS J1507-622.
Unfortunately, this observation was heavily affected by the occurrence of a huge soft proton flare, which led to a good time interval (GTI) of only 0.8 ks for the PN detector, 8.0 ks for MOS1, and 9.2 ks for MOS2 detector. The soft proton flare was identified by extracting the high-energy ($>$ 10 keV for MOS and $10-12$ keV for PN) light curve of the whole observation and was confirmed by the radiation monitor onboard {\it XMM-Newton}. 
However, because of the large intensity of the soft proton flare, residual background proton events still affect the remaining GTI, as they lead to overcorrection of the vignetting, in particular in the outer regions of the PN and MOS detectors.

With the source detection algorithm {\tt edetect\_chain}, one point-like source (source X3 in Fig. \ref{chandra}) was discovered at the position of $\alpha_{\rm{J2000}} = 15^{\rm{h}} 07^{\rm{m}} 08.9^{\rm{s}} \pm 0.5^{\rm{s}}$, $\delta_{\rm{J2000}} = -62^\circ 16' 44.4'' \pm 0.5''$. 
The spectrum of this source (binning of at least 15 counts per bin) can be described by a power law with photon index of $\Gamma = 4.1 \pm 0.6 $ ($\chi^2$/dof = 8.5/8) taking into account the Galactic absorption of $N_{\rm{H}} = 4.4 \times 10^{21} \;\rm{cm^{-2}}$ (LAB Survey, \cite{Kalberla2005}), which results in a flux of $F_{\rm{2-10keV}} = (1.5 \pm  0.9)  \times 10^{-14} \; \rm{erg \; cm^{-2} \; s^{-1}}$. Another possible description is a thermal plasma model using the {\tt mekal} model of {\tt xspec}, taking into account the Galactic absorption and using a metal abundance of $30\%$ of the abundance of the Sun, which results in $ 0.6 \pm 0.2\; \rm{kT}$ ($\chi^2$/dof = 7.4/8) and a flux of $F_{\rm{2-10keV}} = (6 \pm  2)  \times 10^{-15} \; \rm{erg \; cm^{-2} \; s^{-1}}$.
The value of $ 0.6 \pm 0.2\; \rm{kT}$  for the thermal model is compatible with studies on stars that resulted in reasonable values of $ 0.3 - 0.8 \; \rm{kT}$ \cite{Favata2004}.
Because of the low statistics it cannot be determined which model would be preferred. 
\\


This source is identified as a star (TYC 9024.1705.1) of the Tycho catalog (\cite{tycho2}) with a magnitude of $11.70\;\rm{mag}$ in the VT and $12.04\;\rm{mag}$ in the BT band, the corresponding Johnson photometry would be $11.67\;\rm{mag}$ in the optical V band, following calculations by \cite{tycho2}.

This source is very bright in the OM optical and UV filter bands (see Table 1) with $11.96\pm 0.01\;\rm{mag}$ in the U-band. 





\subsection{{\it Chandra} observations in 2009}

The 20 ks observation with {\it Chandra} in 2009 was partially reported in \cite{1507} and is summarized in Fig. \ref{chandra}, showing the smoothed and background-subtracted count map between 0.3 and 8 keV.

With the source-detection algorithm {\tt celldetect}, eleven point-like sources (mainly identified as stars) and one extended source were discovered.

\subsubsection{CXOU J150706.0-621443}

The algorithm {\tt vtpdetect} resulted in the discovery of an additional faint extended source, CXOU J150706.0-621443, (see Fig.\ref{zoomextended}) at the positon $\alpha_{\rm{J2000}} = 15^{\rm{h}} 07^{\rm{m}} 06^{\rm{s}} \pm 3^{\rm{s}}$, $\delta_{\rm{J2000}} = -62^\circ 14' 44'' \pm 1.4''$
that escaped {\tt celldetect}. The algorithm determined 116 counts over a background level of 32 counts for this faint source. 
Since the source is so faint, the tool {\tt specextract} was used to extract a spectrum 
from a circular region of radius $50''$ and an appropriate background region of same size. 
The spectrum is well fit ($\chi^2$/dof = 15.3/11) by a power law of photon index $\Gamma = 1.8^{+ 1}_{-0.8}$ taking into account the Galactic absorption of $N_{\rm{H}} = 4.4 \times 10^{21} \;\rm{cm^{-2}}$ (LAB Survey, \cite{Kalberla2005}). The fit results in a flux of $F_{\rm{2-10keV}} = (6 \pm 3) \times 10^{-14} \;\rm{erg \; cm^{-2} \; s^{-1}}$ in the energy range 2 to 10 keV. 
Even though there unfortunately we were not able to make strong conclusion because of the faintness of the source, it was tentatively interpreted as the possible X-ray PWN remnant of the larger gamma-ray PWN still bright in IC and visible in $\gamma$-rays (\cite{d08}; \cite{myicrc}; \cite{gamma12}). 




\subsubsection{1RXS J150841.2-621006}

The source CXOU J150850.6-621018 can likely be identified as 1RXS J150841.2-621006, because of the known systematics in the positional accuracy of ROSAT (\cite{ROSATsys}). The source detected with {\it Chandra} is located at  
$\alpha_{\rm{J2000}} = 15^{\rm{h}} 08^{\rm{m}} 50.6^{\rm{s}}$, $\delta_{\rm{J2000}} = -62^\circ 10' 18.2''$ 
and the source detection algorithm found an extension of $\sim 36''$. The ROSAT source 1RXS J150841.2-621006 is located at 
 $\alpha_{\rm{J2000}} = 15^{\rm{h}} 08^{\rm{m}} 41.2^{\rm{s}}$, $\delta_{\rm{J2000}} = -62^\circ 10' 06.5''$ 
 with statistical positional uncertainty of $17''$, as reported in \cite{faintRASS}.

\begin{figure}
\centering
\includegraphics[width=\columnwidth]{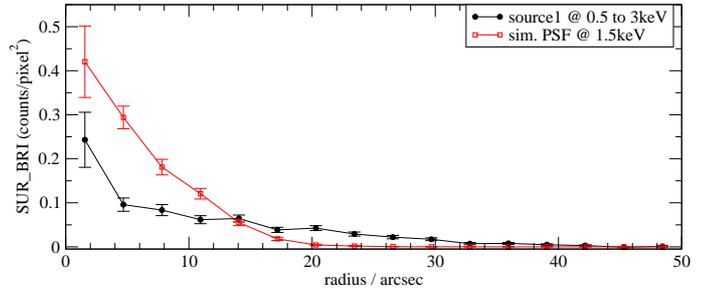}
\caption{Radial profile of the Chandra source CXOU J150850.6-621018 obtained from the energy range 0.5 to 3 keV shown as black circles. We plot with red open squares the radial profile of the PSF at an energy of 1.5 keV.
}
\label{radprofile_chandra}
\end{figure}

Initially (\cite{1507old}), 1RXS J150841.2-621006 (\cite{faintRASS}) was thought to be a pulsar powering a TeV-bright offset PWN, which is common (e.g. \cite{aharonian06}), but this senario now seems unlikely because the source has been found to have an extension, as shown in Fig. \ref{radprofile_chandra}.
To be clearer on this point, if the off-set bright source were a naked pulsar, it might indeed be a classical offset TeV PWN. 
If instead it is a pulsar with enough spin-down luminosity to power a bright and extended PWN, it cannot avoid the inverse Compton scattering from those electrons and therefore also the possible TeV PWN has to extend in the direction of the possible X-ray PWN. This is not the case because the CXOU J150850.6-621018/XMMU J150851.1-621017 is well outside the 3 sigma significance contours of HESS J1507-622.
It might still be that 1RXS J150841.2-621006 is a very young PWN (not correlated to HESS J1507-622) that is not old enough to have been able to accumulate enough leptons to be visible in TeV gamma-rays (e.g. \cite{Vost}).


To identify the real extension of the source, the PSF for this specific off-axis angle and the energy 1.5keV was derived and compared with the radial profile of the source in the energy range 0.5 to 3 keV. 
As can be seen in Fig. \ref{radprofile_chandra}, the detected X-ray source is extended beyond the PSF up to 30-40 arcsec. 

\begin{figure}
\includegraphics[width=0.8\columnwidth]{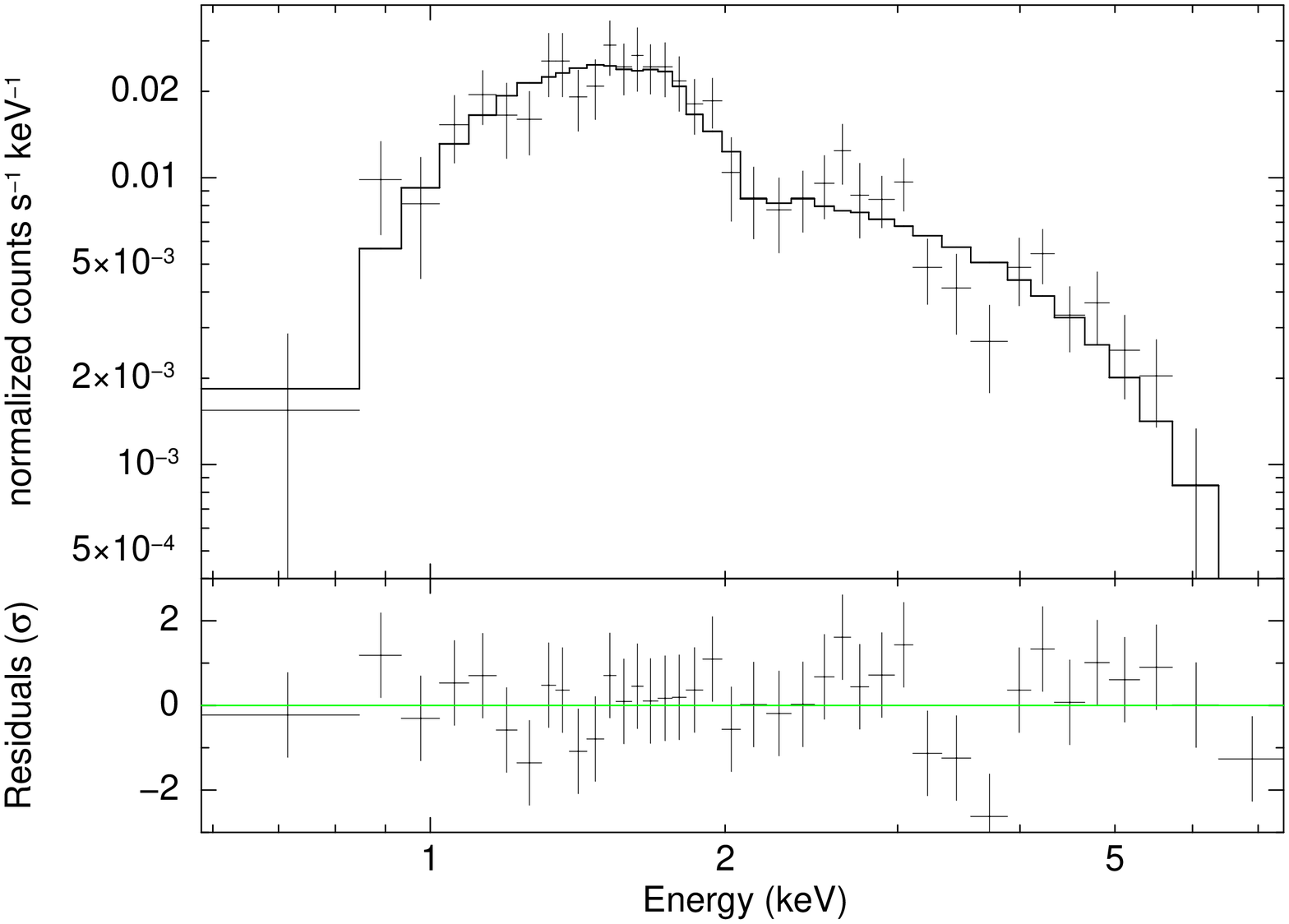}
\vfil
\includegraphics[width=0.8\columnwidth]{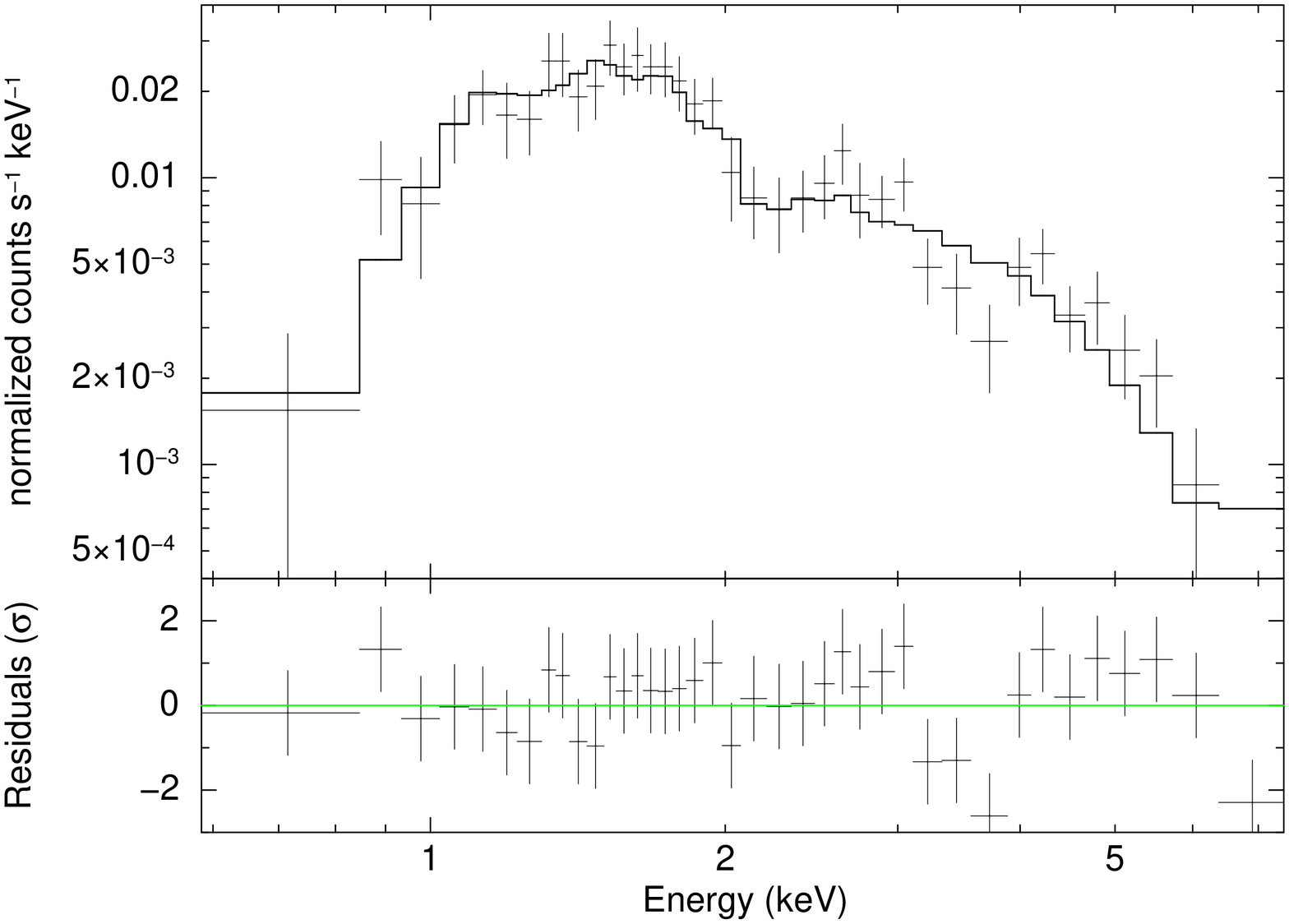}
\caption{{\it Top}: spectrum of CXOU J150850.6-621018 obtained from the {\it Chandra} observation fit with a simple power law taking into account photoelectric absorption.
{\it Bottom}: spectrum of CXOU J150850.6-621018 obtained from the {\it Chandra} observation fit with a thermal model {\tt mekal} taking into account photoelectric absorption. 
}
\label{spec1}
\end{figure}

The spectrum was extracted from a circular region with radius of $\sim 50''$ around the source, and an appropriate region for the background with the same radius was used. The spectrum was binned such that there are at least 25 counts per bin. 
The spectrum can be well fit ($\chi^2$/d.o.f. = 30.4/36) by a simple power law (Fig. \ref{spec1}) 
with spectral index $= 2.0 \pm 0.3$ and an absorption of $N_{\rm{H}} =
(8 \pm 3) \times 10^{21}$ cm$^{-2}$. This could indicate that 1RXS
J150841.2-621006 is a PWN (but not related with HESS J1507-622, since
it is well outside the 3$\sigma$ significance contours).
But the spectrum is equally well described by a thermal plasma
(model {\tt mekal} in {\tt xspec}) with kT $= 5^{+ 5}_{- 2}$ keV and
$N_{\rm{H}} = (7 \pm 3) \times 10^{21}$ cm$^{-2}$  ($\chi^2$/d.o.f. =
34.7/36);  this is is generally indicative of a thermal shell of a supernovae remnant (SNR).
Both fits result in a flux of $F_{\rm{2-10keV}} = (7 \pm 1) \times 10^{-13} \;\rm{erg \; cm^{-2} \; s^{-1}}$ in the energy range 2 to 10 keV.
The fit with higher absorption values than the Galactic absorption yields a better fit than with the Galactic absorption as a fixed value, but the additional amount of absorption has a large uncertainty. \\


\begin{figure}
\centering
\includegraphics[width=\columnwidth]{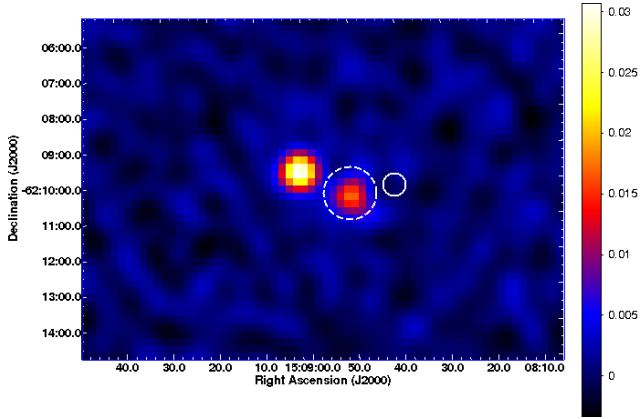}
\caption{Molonglo Galatic Plane Survey (\cite{molonglo}) image around 1RXS J150841.2-621006. The white dashed circle with a radius of $40''$ represents the source CXOU J150850.6-621018 detected by {\it Chandra} and the white circle the position of 1RXS J150841.2-621006 as presented in the RASS (\cite{faintRASS}).
The source MGPS J150850-621025 is spatially consistent with CXOU J150850.6-621018.
}
\label{x1radio}
\end{figure}

The source is also visible at radio wavelengths in the Molonglo Galatic Plane Survey (\cite{molonglo}): the source MGPS J150850-621025 is spatially coincident with 1RXS J150841.2-621006 and the two sources have the same angular extension, as shown in Fig. \ref{x1radio}. The radio flux density of the source MGPS J150850-621025 is $(22.9 \pm 1.4)\; \rm{mJy}$ at 843 MHz (\cite{molonglo}).

However, even when also considering the radio spectral slope in a
combined fit, its spectral energy distribution does not help to distinguish between the two possibilities (PWN or SNR). 

\subsection{{\it XMM-Newton} observation in 2011}

The 39 ks observation with {\it XMM-Newton} on March 2, 2011 was unfortunately also affected by the occurrence of a large soft proton flare detected in the high energy light curves.
The recommended cuts on the high energy light curves lead to a good time interval (GTI) of 1 ks for the PN detector, 4.2 ks for MOS1 and 5.1 ks for MOS2 detector.

Interestingly, only a very faint flare in the Low Energy Detector (sensitive for protons (E$>$1.3 MeV) and electrons (E$>$130 keV)) of the radiation monitor that was not significant in the High Energy Detector  (particle detector sensitive to electrons $>$1 MeV  and protons with energies $>$8 MeV)\footnote{http://xmm2.esac.esa.int/external/xmm\_obs\_info/radmon/ \\ /radmon\_details/index.php} was detected during this soft proton flare.
The detected X-ray flare can be interpreted as a proton or electron flare from the Sun of lower energies ($<$ 130 keV) that resulted in very high count rates in the {\it XMM-Newton} EPIC detectors.
Conversely, in the {\it XMM-Newton} observation in 2009 the flare seen in the radiation monitor was so high that it passed the upper limit for safe operation of the instruments, and the observation was stopped.




With the source-detection algorithm {\tt edetect\_chain}, 
seven sources were found in the MOS2 detector, which are listed in
Table \ref{XMM_sources}. For the PN detector with the lower exposure
only the first five sources were detected. The {\tt ewavelet} tool, which is better for detecting faint sources, was also applied and resulted in the detection of nine sources at a level of at least $5\sigma$. The two sources in addition to those previously detected are named X8 and X9 in the table.
Most of the detected sources are spatially coincident with stars from the Tycho (\cite{tycho2}) and Guide Star Catalog (\cite{gsc}).

From the nine detected sources only four are within the field of view of the OM detector, and the measured optical and UV brightness is given in Tables 1 and 2. Of these four sources, one is bright in all OM filters, identified as star TYC 9024.1705.1.  The other sources are too faint for detection in the UV filter. 


The difference of the 2011 observation with that of 2009 is demonstrated by
analyzing the data of the Radiation Monitor onboard {\it XMM-Newton}
and by noting that the intensity of the soft proton flare is more
than one order of magnitude lower than that of the 2009 data.

\subsubsection{XMMU J150851.1-621017}

\begin{figure}
\centering
\includegraphics[width=\columnwidth]{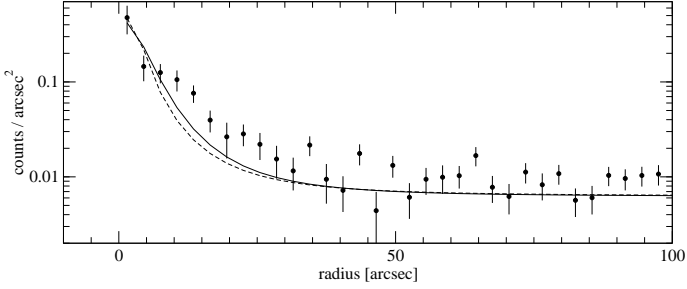}
\caption{Radial profile of the source XMMU J150851.1-621017 (X1) from the {\it XMM-Newton} MOS2 observation obtained with {\tt eradial} for a binsize of $3''$. The lines present the fit of a King profile to the radial profile for the monochromatic energies 0.5 keV (solid) and 10 keV (dashed line).}
\label{XMM_X1_radprofile}
\end{figure}

For the source XMMU J150851.1-621017 (X1), which can possibly be identified as 1RXS J150841.2-621006 in the previous {\it Chandra} observation, a radial profile with a fit of a King profile for the energies 0.5 and 10 keV was produced.
The radial profile is shown in Fig. \ref{XMM_X1_radprofile} and confirms the result from the previous {\it Chandra} observation that the source is slightly extended over the PSF of the instrument. 

The light curve of this source does not show significant variation, but because of the cut for the soft proton flare, a low exposure in the GTI and hence low statistics resulted.

The spectrum was extracted from a circular region with radius $50''$ and an appropriate background on the same chip was chosen using the same region size. The spectrum was binned to at least 15 counts/bin because of the low statistics. It can be fit with a power law with photon index of $\Gamma = 1.5 \pm 0.4$ taking into account the Galactic absorption of $N_{\rm{H}} = 4.4 \times 10^{21} \;\rm{cm^{-2}}$ (LAB Survey, \cite{Kalberla2005}). The goodness of fit is $\chi^2/dof = 6.6/9$ because of the low statistics. To leave the parameter of the absorption as free parameter does not result in a reasonable fit, so that the amount of absorption determined from the {\it Chandra} observation cannot be verified. 
The flux in the energy range 2 to 10 keV is $F_{\rm{2-10keV}} = (9 \pm 2) \times 10^{-13} \;\rm{erg \; cm^{-2} \; s^{-1}}$, perfectly compatible with {\it Chandra} results. Note that the value is the same as the flux listed in Table 1 and calculated in the 0.2-12 keV band with {\tt edetect\_chain}, which converts the source count rate into flux values depending on fixed conversion factors, assuming a specific spectral index and absorption column density.



\subsubsection{XMMU J150835.7-621021}

Interestingly, one source was not detected in the {\it Chandra} observation; XMMU J150835.7-621021 (X2), which is clearly detected with {\it XMM-Newton}. 
Its radial profile was detemined to identify its extension and to decide whether it is spatially connected to XMMU J150851.1-621017 (X1). The radial profile was well fit with a King profile for the energies 0.5 and 10 keV, which is representative for the PSF of the instrument. The sources X1 and X2 are located outside of the FoV of the first XMM observation and their results therefore cannot been compared with the 2009 observation.
This is why we found no significant extension or spatial connection to XMMU J150851.1-621017 (X1).


A possible counterpart for the source XMMU J150835.7-621021 (X2) was found in the Guide Star Catalog (\citep{gsc}). 
The sources GSC2.3 S7QP006140 and GSC 2.3 S7QP006099 are both spatially consistent with X2 ($7''$ and $8''$ away from X-ray position, respectively). The brightness in the GSC is given in the bands F (in the ragne 600-750 nm), V (between 500 and 600 nm) in the optical and N (between 8 and 15 micron) in infrared for these two sources. 
GSC2.3 S7QP006140 has a brightness of F = 15.19 mag, V = 16.28 mag and  N = 14.11 mag and 
GSC 2.3 S7QP006099 has a brightness of F = 16.04 mag, V = 16.79 mag, N = 15.45mag (\citep{gsc}).

The light curve of this source does not show significant variation,
but again because of the cut for the soft proton flare, a low exposure in the GTI and hence low statistics resulted.

The spectrum was similarly extracted from a circular region with radius $30''$ and an appropriate background on the same chip was chosen with a radius of $50''$. The spectrum was binned with 10 counts/bin because of the low statistics. It can be fit with a power law with photon index of $\Gamma = 2.6 \pm 0.9$ taking into account the Galactic absorption of $N_{\rm{H}} = 4.4 \times 10^{21} \;\rm{cm^{-2}}$. The goodness of fit is $\chi^2/dof = 3.8/3$ because of the low statistics. 
The flux in the energy range 2 to 10 keV is $F_{\rm{2-10keV}} = (8 \pm 4) \times 10^{-14} \;\rm{erg \; cm^{-2} \; s^{-1}}$. 

The determined flux is above the point-source sensitivity limit\footnote{http://cxc.cfa.harvard.edu/cdo/about\_chandra/} of {\it Chandra} and should have been detected in the previous {\it Chandra} observation. Therefore we conclude that the source is variable. We searched the {\it Chandra} observation at this position for a signal. A radial profile at the position of XMMU J150835.7-621021 (X2) was created and resulted in a surface brightness comparable to the background level and hence was not detected. 



A simple explanation for this source might be a background active galactic nucleus (AGN) or an X-ray binary (XRB). But the AGN does not seem to be supported by observations: at this position there is no AGN or any possible AGN hosting galaxy revealed at any wavelength (e.g. Fig. \ref{x1radio}).

An XRB scenario that involves the star GSC2.3 S7QP048279 might be a viable option; but, to support this scenario, we would need to know the spectral type of the star GSC2.3 S7QP048279 (the second-generation guide star catalog (\cite{gsc}) does not have detailed information). A high-mass X-ray binary (HXRB) would require a massive star, while the low X-ray flux measured with {\it XMM-Newton} is perfectly compatible with a low mass X-ray binary (LXRB) scenario. 

But there is another option to explain this variable source: GSC2.3 S7QP048279 might be a flaring star. This phenomenon normally occurs on time scales from seconds to days, generally with a fast rise and exponential-decay light curve.
Flaring stars generally show a strong increase in flux both in the optical and in the X-ray bands, and flares are expected from stars of several spectral types and especially from M-dwarf stars such as CN Leonis (\cite{liefke}). Unfortunately the low statistics on the light curve and the fact that OM does not cover this region of the sky do not help in confirming or disproving this scenario.
We note that the star GSC2.3 S7QP048279, consistent with the X-ray source, has a similar brightess in the optical as CN Leonis. For CN Leonis, the brightness of the source increased by a factor of 100 during a strong flare. Such amplitude of variation might explain the nondetection in 2009 with Chandra during quiescent state of the star. Unfortunatly, the second-generation guide star catalog (\cite{gsc}) does not provide spectral information.
But a CN Leonis flaring star remains a viable option to explain XMMU J150835.7-621021, and this appears to be the most likely solution.






\begin{figure}
\centerline{
\includegraphics[width=0.45\columnwidth]{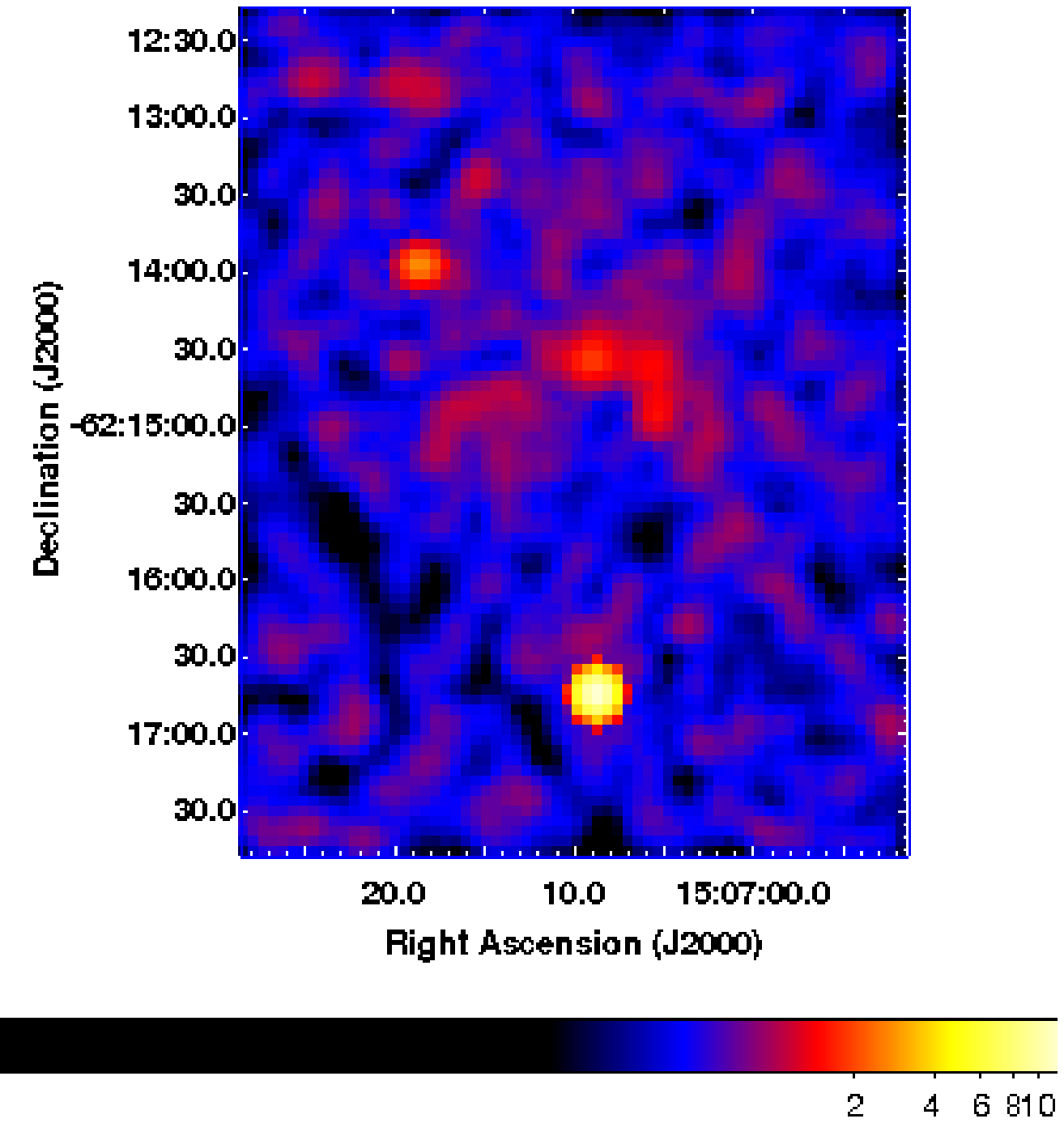}
\hfil
\includegraphics[width=0.45\columnwidth]{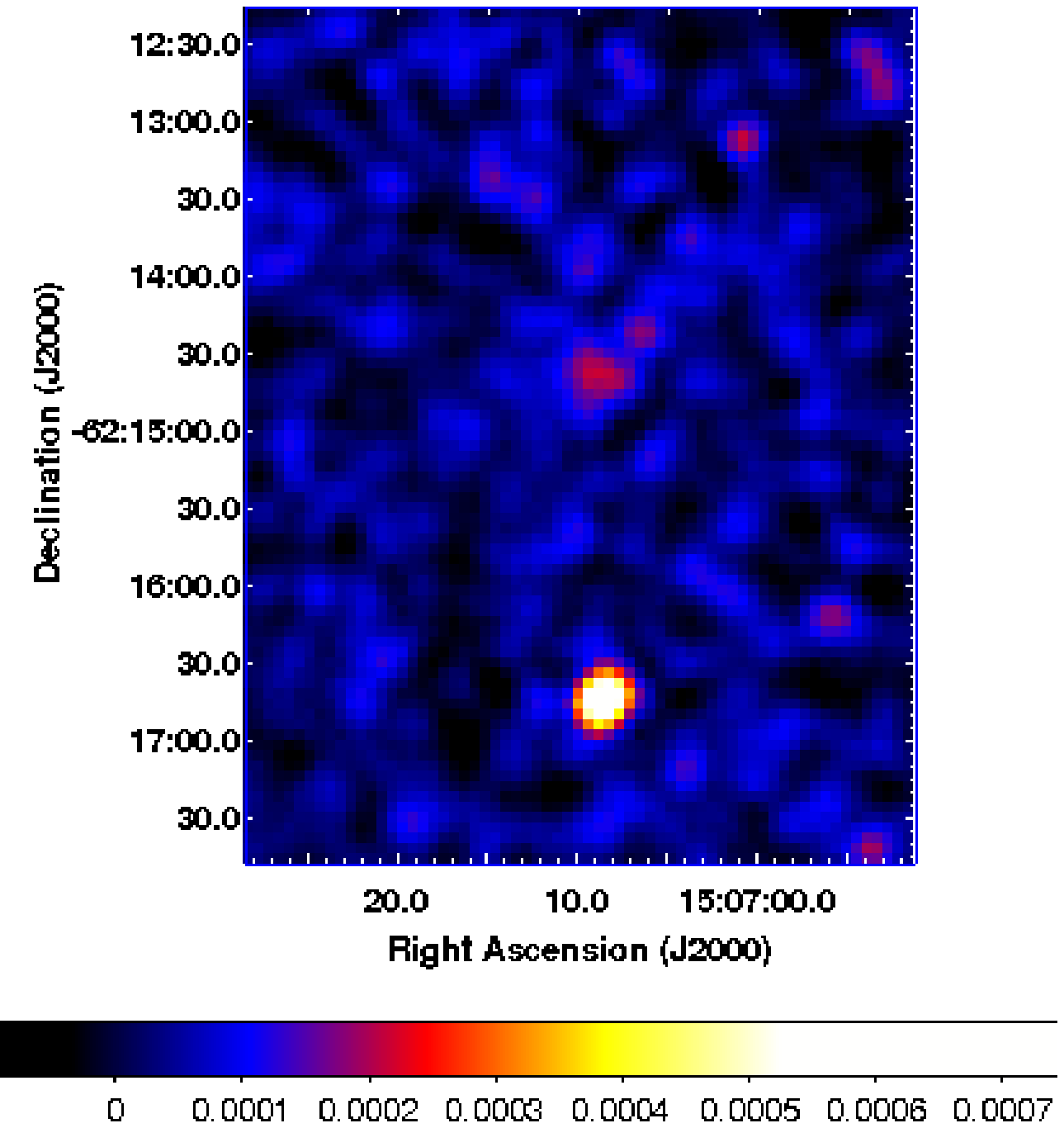}
}
\caption{{\it Left}: zoom-in of the exposure-corrected smoothed image of the {\it Chandra} observation at the position of the faint extended source.
{\it Right}: zoom-in of the background- and exposure-corrected mosaic of the MOS  detector images at the position of the faint extended source. This image has been smoothed with a Gaussian with radius of 3 pixels.
}
\label{zoomextended}
\end{figure}

\subsubsection{Faint extended source}


The faint extended source CXOU J150706.0-621443 detected with {\it
  Chandra} that is spatially coincident with HESS J1507-622 is also detected
with this {\it XMM-Newton} observation. Because of the soft proton flare
and the consequently lower exposure in the GTI, the extended source is
faintly visible in the exposure-corrected mosaic of the MOS detectors,
as shown in Fig. \ref{zoomextended}. With the source-detection algorithm {\tt ewavelet} using a threshold limit of 4
$\sigma$, a significant detection can be made, however. Because of the lower exposure, the detected extension is lower than the extension detected with the {\it Chandra} observation. 
But it remains the only possible X-ray counterpart of HESS J1507-622.

\section{Summary and conclusions}

HESSJ1507-622 was observed several times with XMM-Newton and Chandra
to identify its nature. Because of the location $3.5^\circ$ away from the
Galactic plane, the photoelectric absorption of X-rays is much lower
than in the Galactic plane which makes the search for X-ray
counterpart of this unidentified very high energy source very promising.

The XMM-Newton and Chandra observations revealed several new
point-like X-ray sources, of which most have stars as counterparts.
Two extended X-ray sources were discovered. One of them is
associated with a Rosat source and is located outside of the VHE
extension of HESS J1507-622, which makes an association unlikely.
The second extended X-ray source is very faint with a flux of $\sim
6\times 10^{-14} \;\rm{erg \; cm^{-2} \; s^{-1}}$ and can be
interpreted as the possible X-ray PWN remnant of the larger gamma-ray
PWN that is still bright in inverse-Compton emission and visible in
gamma-rays.
This faint extended source is confirmed with the XMM-Newton
observation in 2011, although because of the faintness of the source and
because the observation was affected by a soft proton flare that dramatically reduced
the GTI, the source appears very faint and less extended.

The deep X-ray observations only partially shed light on HESS J1507-622; the faint extended source detected with {\it Chandra} (that is partially confirmed by the {\it XMM-Newton} hints of detection) is still interpreted as the possible X-ray remnant of the TeV PWN that is visible as HESS J1507-622, and it remains the only possible counterpart for it; for pulsars older than $\sim 10^3$ years the VHE PWNe are typically 100-1000 times larger than the sizes of the X-ray PWNe (while the difference is only a factor 2 for some younger PWNe, like the Crab nebula), as shown by \cite{kar}.
This is important to constrain the age of the unidentified VHE source: as shown by \cite{myicrc}, the sringent upper limit provided by the Molonglo Galactic Plane Survey is especially constraining and implies that HESS J1507-622 is more than 20 kyr old.

This scenario has several caveats that were already discussed by \cite{1507}, however. There is no pulsed emission detected in the HESS J1507-622 region; on the other hand, it is expected that for an ancient PWN the pulsar might have spun-down below the sensitivity of the current generation of instruments. If the age of the unidentified source might be close to the calculated UL of 50 kyr, the pulsar might still be visible in X-rays (i.e., is young enough to have a hot thermally emitting surface that should be visible in X-rays from any direction), but even then the pulsar emission cone might be outside our line of sight. Another possibility is that the velocity kick (e.g., as in the case described by \cite{aharonian06}) might have pushed the pulsar well outside the VHE emission region (e.g., in this latter case the progenitor pulsar might be 1RXS J150639.1-615704 or 1RXS J150354.7-620408, visible in Fig. 1 of \cite{1507}: but currently there are not enough observations to prove or disprove them as possible counterparts).
Alternatively, the pulsar might be an anomalous X-ray pulsar and therefore spinning down much quicker than expected for a normal pulsar (such as CXO J184454.6-025653 discussed by \cite{12}). However, fot this latter case we note that the X-ray emission from an anomalous X-ray pulsar is not thought to be powered by rotation. The conventional understanding is that its X-ray emission is powered by magnetic field decay, and so it is largely independent of spin parameters. While there are transient magnetars that are very weak or undetectable in their quiescent state, this may be because they are very old and their magnetic field has already decayed. Indeed, some of these transients  have small dipole fields as measured by their spin-down rates. But the problem with magnetars powering unidentified TeV sources is the same as the problem with ordinary pulsars powering unidentified TeV sources: even though their X-ray emission mechanisms are different, they might both be still bright enough to detect in X-rays.

Two other interesting sources were discussed, even though they cannot be plausible counterparts for HESS J1507-622. CXOU J150850.6-621018 was confirmed to be an extended source by {\it XMM-Newton} (XMMU J150851.1-621017) and a plausible counterpart of the ROSAT source 1RXS J150841.2-621006. 
CXOU J150850.6-621018 is most likely a newly discovered SNR or PWN: unfortunately, the spectral slope in X-rays and in radio is compatible with both senarios and did not help in indentifying the exact nature of this source.
However, given its extension and its location (outside the 3$\sigma$ significance contours of HESS J1507-622), it is confirmed that it cannot be an alternative possible counterpart of HESS J1507-622.

Another newly discovered source is a variable X-ray source that was
detected in the XMM-Newton observation of 2011 and was not
significantly detected in 2009.
This variable X-ray source (XMMU J150835.7-621021) detected in the
XMM-Newton observation of 2011 is spatially consistent with a star of
the GSC catalog S7QP048279. Its variable behavior (detection in 2011
and nondetection in 2009) may be interpreted as belonging to a flaring star.
On the short-time light curve, unfortunately only few counts remain in
the GTI, which together with the faintness of the source result in
large uncertainties. Therefore the typical behavior of a flaring star
with a steep short increase and an exponential decay cannot be
confirmed.

\newpage
\begin{table*}
\begin{center}
\begin{tabular}{|cccccccc|}
\hline
src	& XMMU name		& RA (J2000)	& Dec(J2000)	& $\Delta$ RA,DEC & $F_{\rm{0.2-12 keV}}/10^{-13}$	& extension & CXOU name\\
& & &	& [arcsec] & [$\rm{erg \; cm^{-1} \; s^{-1}}$] & [arcsec] & \\
\hline
X1 & XMMU J150851.1-621017	& 15:08:51.12  & -62:10:17.04	& 0.98 & $9 \pm 2$	& $16 \pm 3$ &  CXOU J150850.6-621018 \\
X2 & XMMU J150835.7-621021 & 15:08:35.76 & -62:10:21.00 & 1.6 & $1.5 \pm 0.5$ & $12 \pm 2$ &  \\
& & & & &  &  &  \\
X3 & XMMU J150708.4-621642 & 15:07:08.40 & -62:16:42.24 & 1.3 & $1.6 \pm 0.7$ &  $4 \pm 1$ & CXOU J150708.8-621643\\
X4 & XMMU J150620.8-621113 & 15:06:20.88 & -62:11:13.20 & 2.5 & $0.7 \pm 0.4$ & $12 \pm 2$& CXOU J150621.7-621110\\
X5 & XMMU J150802.8-621756 & 15:08:02.88 & -62:17:56.76 & 2.1 & $0.6 \pm 0.4$ & $4 \pm 1$ & \\
X6 & XMMU J150755.9-622237 & 15:07:55.92 & -62:22:37.92 & 1.5 & $2.3 \pm 0.8$  & $2 \pm 0.6$ & CXOU J150756.0-622238\\
X7 & XMMU J150728.0-622611 & 15:07:28.08 & -62:26:11.76 & 1.5 & $2 \pm 1$  & $2.9 \pm 0.7$ & \\
\hline
src	& XMMU name		& RA (J2000)	& Dec(J2000)	& $\Delta$ RA,DEC & counts/sec & extension & CXOU name\\
& & 	& & [arcsec] &  & [arcsec] &\\
\hline
X8 & XMMU J150832.1-620744 & 15:08:32.16 & -62:07:44.40 & 0.8 & $(12 \pm 4)$   & $5 \pm 1$ & \\   
X9 & XMMU J150916.3-621658 & 15:09:16.32 & -62:16:58.80 & 2.8 & $(6 \pm 3)$  & $2 \pm 0.4$ & \\
\hline
\end{tabular}
\end{center}
\label{XMM_sources}
\caption{Summary of the seven sources detected by {\it XMM-Newton} on the MOS2 detector using {\tt edetect\_chain}. The RA and DEC uncertainty is calculated as the square of the quadratic sum of the errors in RA and DEC and corresponds to $\sim$1 standard deviation uncertainty.
The flux was calculated with {\tt edetect\_chain} using the conversion factors mentioned in the text. 
For source X2, the flux calculated directly from the detailed spectral analysis is given instead, since it is very faint and the photon index of its spectrum is 2.6.
The sources X8 and X9 were additionally detected with {\tt ewavelet}. 
The extension mentioned for all sources results from the tool {\tt ewavelet}, even if the radial profiles of some faint off-axis sources (such as X2) are compatible with a King profile.  
The CXO names of the sources detected with {\it Chandra} are given as well. 
%
}
\end{table*}

\begin{table*}
\begin{center}
\begin{tabular}{|cccccccc|}
\hline
src	& UVW1 & UVW2 & UVM2 & U & B & V & poss. ident.	\\
 & mag & mag & mag & mag & mag & mag & \\
\hline
X1 &x &x &x &x &x &x & 1RXS J150841.2-621006 \\
X2 &x &x &x &x &x &x & 1RXS J150841.2-621006 \\
& & & & & & & GSC2.3 S7QP006140 \\
& & & & & & & GSC2.3 S7QP006099 \\
X3 & $12.4082 \pm 0.0003$ & $14.22 \pm 0.02$ & $13.942 \pm 0.003$ & $11.96 \pm 0.01$ & $\pm$ & $11.39 \pm 0.03$ & TYC 9024.1705.1  \\
X4 &x &x &x &x &x &x   & TYC 9024.1615.1  \\
X5 &- &- &- &- & $20.7 \pm 0.2$ & $18.8 \pm 0.1$  & GSC2.3 S7QP084521 \\
X6 &- &- &- & $21.4 \pm 0.6$ & $20.2 \pm 0.1$ & $18.5 \pm 0.1$  & GSC2.3 S7QP035739  \\
X7 &- &- &- &- & $19.67 \pm 0.09$ & $17.92 \pm 0.06$  & GSC2.3 S7QP002061 \\
\hline
src	& UVW1 & UVM2 & UVW2 & U & B & V & poss. ident.	\\
 & mag & mag & mag & mag & mag & mag & \\
\hline
X8 &x &x &x &x &x &x  & GSC2.3 S7QP007261 \\   
X9 &x &x &x &x &x &x  & GSC2.3 S7QP084895 \\
\hline
\end{tabular}
\end{center}
\label{XMM_sources2}
\caption{Summary of the sources detected by {\it XMM-Newton}. 
The brightness detected with the optical monitor (OM) is given when the source was in the field of view of this detector and bright enough for a detection. x marks the sources that are outside of the FoV of the OM detector.
For source 1 only the probable identification 1RXS J150841.2-621006 (\cite{faintRASS}) is mentioned (the possible identification with the ROSAT source is underlined by the flux of CXOU J150850.6-621018/XMMU J150851.1-621017 and its spectral shape). The Tycho-2 catalog (\cite{tycho2}) and the second-generation guide star catalog (\cite{gsc}) have been inspected when searching for possible counterparts. The search for reasonable counterparts included not only the three standard deviations uncertainty in the X-ray position (which is determined with a PSF fit), but also the extension of the source. Considering the systematics (\cite{ROSATsys}), X2 is spatially also compatible with the ROSAT source 1RXS J150841.2-621006.
}
\end{table*}

\newpage

\begin{acknowledgements}

The authors would like to thank again the other members of the ''H.E.S.S. 1507 team`` and its referees: W. Hofmann, W. Domainko, N. Komin, O. Reimer, S. Wagner, and most of all, O. de Jager. In particular, we dedicate this paper to the memory of O. de Jager, who initiated and implemented the first models of ancient PWNe; he is still a great inspiration to us as a scientist and a wonderful human being. Okkie, learning from you was a great pleasure and working with you was a great honor: we will miss you very much!
We also acknowledge K. Mannheim, D. Els\"asser and C. Liefke for the useful discussions.
Finally, we would like to acknowledge the anonymous A\&A referee for the constructive feedback in the referee process and for his/her work to improve the paper.


\end{acknowledgements}


\begin{thebibliography}{}


%
%
%
\bibitem[Aharonian et al. 2006a]{survey2} Aharonian F. et al. (H.E.S.S. Collaboration)
2006, ApJ, 636, 777

%
%
\bibitem[Aharonian et al. 2006b]{aharonian06} Aharonian F. et al. (H.E.S.S. Collaboration) 2006, A\&A, 460, 365

   \bibitem[Aharonian et al. 2008]{unid} Aharonian F. et al. (H.E.S.S. Collaboration) 2008,
      A\&A, 477, 353



%
%
%
%
%
%
%
%
%
   \bibitem[de Jager 2008]{d08} de Jager O.C. 2008, ApJ, 678, L113
%
%
%
%

   \bibitem[Favata et al. 2004]{Favata2004} Favata F. et al. A\&A 2004, 418, L13.
%
   \bibitem[Ferreira \& de Jager 2008]{fd08} Ferreira S.E.S. and de Jager O.C. 2008, A\&A, 478, 17
%
%
%
%
%
   \bibitem[Green et al. 1999]{molonglo} Green A.J. et al. 1999, ApJS, 122, 207
%
%
%
   \bibitem[Harris et al. 1998]{ROSATsys} Harris D. E. et al. 1998, A\&AS, 133, 431
%
%

\bibitem[H.E.S.S. collaboration 2011]{1507}H.E.S.S. collaboration 2011, A\&A, 525, id.A45

   \bibitem[Hog et al. 2000]{tycho2} Hog E. et al. 2000, A\&A, 335, L27
%
%
   \bibitem[Kalberla et al. 2005]{Kalberla2005} 	
	Kalberla P. M. W. et al. 2005, A\&A, 440, 775
%
   \bibitem[Kargaltsev \& Pavlov 2010]{kar} Kargaltsev O. and Pavlov P. O. 2010, AIP Conference Series, 1248, 25
%
%
%
  \bibitem[Lasker et al. 2008]{gsc} Lasker B. M. 2008, AJ, 136, 735

\bibitem[Liefke et al. 2010]{liefke} Liefke C., Fuhrmeister B. and Schmitt J. H. M. M.  2010, A\&A, 514, 94

%
%

\bibitem[Mason et al. 2001] {Mason2001} Mason K.~O. et al. 2001, A\&A, 365, L36
%

\bibitem[Nolan et al. 2012]{2FGL} Nolan P.~L. et al. (\emph{Fermi LAT} collaboration) 2012, ApJSS, 199, 31



%
%
%
%
%

\bibitem[Tam et al. 2006]{12} Tam C. R. et al. 2006, ApJ, 652, 548.

%
%
%

\bibitem[Tibolla et al. 2009]{1507old}  Tibolla O. et al. 2009, arXiv:0912.4229

\bibitem[Tibolla et al. 2011]{myicrc} Tibolla O. et al. 2011, arXiv:1109.3144.

\bibitem[Tibolla et al. 2012]{gamma12} Tibolla O. et al. 2012, AIP conf. proc., 1505, 349; arXiv:1109.3144.
%
%
  \bibitem[Voges et al. 2000]{faintRASS} Voges W. et al. 2000,
      IAU Circ., 7432, 1

\bibitem[Vorster et al. 2013]{Vost} Vorster M. J., Tibolla O., Ferreira S. E. S. and  Kaufmann, S. 2013, ApJ, 773, id. 139.



%
%
%
%
%





\end{thebibliography}
\end{document}